\documentclass[fleqn,twoside]{article}
\usepackage[headings]{espcrc2}

\readRCS
$Id: espcrc2.tex,v 1.2 2004/02/24 11:22:11 spepping Exp $
\usepackage{graphicx, balance}
\usepackage{amsmath}
\usepackage{amssymb}
\usepackage{amsfonts}
\usepackage{tikz}
\usepackage{textcomp}
\usepackage{mathtools}
\usepackage{setspace}
\usepackage{fancyhdr}
\usepackage[figuresright]{rotating}

\title{\textbf{An Epitome of Multi Secret Sharing Schemes for General Access Structure}}

\author{V P Binu\address[DCA]{Department of Computer Applications, Cochin University  of Science and Technology, \\~Cochin-682022 India, Contact: binuvp@gmail.com \\},
A  Sreekumar\addressmark}

\runtitle{An Epitome of Multi Secret Sharing Schemes for General Access Structure}
\runauthor{Binu V P, et al.,}

\begin{document}
\begin{abstract}

Secret sharing schemes are widely used now a days in various applications, which need more security, trust and reliability.\@ In secret sharing scheme, the secret is divided among the  participants and only authorized set of participants can recover the secret by combining their shares.\@ The authorized set of participants are called access structure of the scheme. In Multi-Secret Sharing Scheme (MSSS), $k$ different secrets are distributed among the participants, each one according to an access structure.\@ Multi-secret sharing schemes have been studied extensively by the cryptographic community.\@ Number of schemes are proposed for the threshold multi-secret sharing and multi-secret sharing according to generalized access structure with various features. In this survey we explore the important constructions of multi-secret sharing for the generalized access structure with their merits and demerits. The features like whether shares can be reused, participants can be enrolled or dis-enrolled efficiently, whether shares have to modified in the renewal phase $etc$., are considered for the evaluation. 
\end{abstract}

\maketitle

\section{INTRODUCTION}
Secret sharing schemes are important tool used in security protocols.\@ Originally motivated by the problem of secure key storage by Shamir \cite{shamir1979}, secret sharing schemes have found numerous other applications in cryptography and distributed computing.\@ Threshold cryptography \cite{desmedt1992shared}, access control \cite{naor1998access}, secure multi-party computation \cite{ben1988completeness} \cite{chaum1988multiparty} \cite{cramer2000general}, attribute based encryption \cite{goyal2006attribute} \cite{bethencourt2007ciphertext}, generalized oblivious transfer \cite{tassa2011generalized}   \cite{shankar2008alternative}, visual cryptography  \cite{naor1995visual} $etc.,$ are the significant areas of development using the secret sharing techniques.
\vskip 2mm
In secret sharing, the secret is divided among $n$ participants in such a way that only designated subset of participants can recover the secret, but any subset of participants which is not a designated set cannot recover the secret.\@ A set of participants who can recover the secret is called an \textit{access structure} or \textit{authorized set}, and a set of participants which is not an authorized set is called an \textit{unauthorized set} or \textit{forbidden set}.
The following are the two fundamental requirements of any secret sharing scheme.
\begin{itemize}
\item \textbf{Recoverability:}Authorized subset of participants should be able to recover the secret by pooling their shares.
\item \textbf{Privacy:}Unauthorized subset of participants should not learn any information about the secret.
\end{itemize}

Let $\mathcal{P}=\{P_i|i=1,2,\ldots,n\}$ be the set of participants and the secret be $K$.\@ The set of all secret is represented by $\mathcal{K}$.\@ The set of all shares $S_1,S_2,\ldots,S_n$ is represented by $\mathcal{S}$. The participants set is partitioned into two classes.
\begin{enumerate}
\item The class of authorized sets $\Gamma$ is called the \textit{access structure.}
\item The class of unauthorized sets $\Gamma^c=2^\mathcal{P}\setminus \Gamma$
\end{enumerate}

Let us assume that $\mathcal{P},\mathcal{K},\mathcal{S}$ are all finite sets and there is a probability distribution on $\mathcal{K}$ and $\mathcal{S}$. We use $H(\mathcal{K})$ and $H(\mathcal{S})$ to denote the entropy of $\mathcal{K}$ and $\mathcal{S}$ respectively.
\vskip 2mm
In a secret sharing scheme there is a special participant called \textit{Dealer} $\mathcal{D} \notin \mathcal{P}$, who is trusted by everyone. The dealer chooses a secret $K \in \mathcal{K}$ and the shares $S_1, S_2,\ldots, S_n$ corresponding to the secret is generated. The shares are then distributed privately to the participants through a secure channel.
\vskip 2mm
In the secret reconstruction phase, participants of an access set pool their shares together and recover the secret. Alternatively participants could give their shares to a combiner to perform the computation for them. If an unauthorized set of participants pool their shares they cannot recover the secret. Thus a secret sharing scheme for the access structure $\Gamma$ is the collection of two algorithms:\\
\textbf{Distribution Algorithm}:This algorithm has to be run in a secure environment by a trustworthy party called Dealer. The algorithm uses the function $f$, which for a given secret $K \in \mathcal{K}$ and a participant $P_i \in \mathcal{P}$, assigns a set of shares from the set $\mathcal{S}$ that is $f(K,P_i)=S_i \subseteq \mathcal{S}$ for $i=1,\ldots,n$.$$f:\qquad \mathcal{K} \times \mathcal{P} \implies 2^\mathcal{S}$$
\textbf{Recovery Algorithm}:This algorithm has to be executed collectively by cooperating participants or by the combiner, which can be considered as a process embedded in a tamper proof module and all participants have access to it. The combiner outputs the generated result via secure channels to cooperating participants. The combiner applies the function,  $$g:\mathcal{S}^t \implies \mathcal{K}$$ to calculate the secret. For any authorized set of participants $g(S_1,\ldots,S_t)=K$, if ${P_1,\ldots,P_t} \subseteq \Gamma$. If the group of participant belongs to an unauthorized set, the combiner fails to compute the secret.
\vskip 2mm
A secret sharing scheme is called perfect if for all sets $B$, $ B \subset \mathcal{P}$ and $B \notin \Gamma$, if participants in $B$ pool their shares together they cannot reduce their uncertainty about $S$. That is, $H(K)=H(K\mid\mathcal{S}_B)$, where $\mathcal{S}_B$ denote the collection of shares of the participants in $B$. It is known that for a perfect secret sharing scheme $H(S_i) \geq H(K)$. If $H(S_i) = H(K)$ then the secret sharing scheme is called ideal.
\vskip 2mm
An access structure $\Gamma_1$ is \textit{minimal} if $\Gamma_2 \subset \Gamma_1$ and $\Gamma_2 \in \Gamma$ implies that $\Gamma_2=\Gamma_1$. Only \textit{monotone access structure} is considered for the construction of the scheme in which $\Gamma_1 \in \Gamma$ and $\Gamma_1 \subset \Gamma_2$ implies $\Gamma_2 \in \Gamma$. The collection of minimal access sets uniquely determines the access structure. The access structure is the closure of the minimal access set. The access structure $\Gamma$ in terms of minimal access structure is represented by $\Gamma_{min}(\Gamma_0)$.
\vskip 2mm
For an access structure $\Gamma$, the family of unauthorized sets $\Gamma^c=2^\mathcal{P} \setminus \Gamma$ has the property that,  given an unauthorized set $B \in \Gamma^c$ then any subset $C \subset B$ is also an unauthorized set. An immediate consequence of this property is that for any access structure $\Gamma$, the set of unauthorized sets can be uniquely determined by its \textit{maximal set}. We use $\Gamma^c_{max}$ to denote the representation of $\Gamma^c$ in terms of maximal set.
\vskip 2mm
For all $B \in \Gamma$, if $|B| \ge t$, then the access structure corresponds to  a $(t,n)$ threshold scheme. In the $(t,n)$ threshold scheme $t$ or more participant can reconstruct the secret. Section 2 gives an insight into the threshold secret sharing schemes. Secret sharing schemes realizing the general access structures are mentioned in Section 3. Section 4 explores the various multi secret sharing techniques in the literature. Section 5 is the summary where different schemes are compared for their merits and demerits. Section 6 is the conclusion.

\section{THRESHOLD SECRET SHARING }
\noindent Development of secret sharing scheme started as a solution to the problem of safeguarding cryptographic keys by distributing the key among $n$ participants and $t$ or more of the participants can recover it by pooling their shares. Thus the authorized set is any subset of participants containing more than $t$ members. This scheme is denoted as $(t,n)$ \textit{threshold scheme}.
\vskip 2mm
The notion of a threshold secret sharing scheme is independently proposed by Shamir \cite{shamir1979} and Blakley \cite{blakley1979} in 1979. Since then much work has been put into the investigation of such schemes. Linear constructions were most efficient and widely used. A threshold secret sharing scheme is called \textit{perfect}, if less than $t$ shares give no information about the secret. Shamir's scheme is perfect while Blakley's scheme is non perfect. Both the Blakley's and the Shamir's constructions realize $t$-out-of-$n$ shared secret schemes. However, their constructions are fundamentally different.
\vskip 2mm
Shamir's scheme is based on   polynomial interpolation over a finite field. It uses the fact that we can find a polynomial of degree $t-1$ given $t$ data points. A polynomial $f(x)=\sum_{i=0}^{t-1}a_ix^i$, with $a_0$ is set to the secret value and the coefficients $a_1$ to $a_{t-1}$ are assigned random values in the field is used for secret sharing. The value $f(i)$ is given to the user $i$ as secret share. When $t$ out of $n$ users come together they can reconstruct the polynomial using Lagrange interpolation and hence obtain the secret.
\vskip 2mm
Blakley's secret sharing scheme has a different approach and is based on hyperplane geometry. To implement a $(t,n)$ threshold scheme, each of the $n$ users is given a hyper-plane equation in a $t$ dimensional space over a finite field such that each hyperplane passes through a certain point.\@ The intersection point of these hyperplanes is the secret.\@ When $t$ users come together, they can solve the system of equations to find the secret.
\vskip 2mm
McEliece and Sarwate \cite{mceliece1981sharing} made an observation that Shamir's scheme is closely related to Reed-Solomon codes \cite{reed1960polynomial}. The error correcting capability of this code can be translated into desirable secret sharing properties. Karnin {\it et al.,} \cite{karnin1983} realize threshold schemes using linear codes. Massey \cite{massey1993minimal} introduced the concept of minimal code words and provided that the access structure of a secret sharing scheme based on a $[n,k]$ linear code is determined by the minimal codewords of the dual code.   
\vskip 2mm
Number theoretic concepts are also introduced for threshold secret sharing scheme.\@ The Mingotee scheme  \cite{mignotte1983} is based on modulo arithmetic and \textit{Chinese Remainder Theorem (CRT)}. A special sequence of integers called Mingotte sequence is used here. The shares are generated using this sequence.\@ The secret is reconstructed by solving the set of congruence equation using CRT. The Mingotte's scheme is not perfect.\@ A perfect scheme based on CRT is  proposed by Asmuth and Bloom \cite{asmuth1983}. They also uses a special sequence of pairwise coprime positive integers. 
   \vskip 2mm    
Kothari \cite{kothari1985generalized} gave a generalized threshold scheme. A secret is represented by a scalar and a linear variety is chosen to conceal the secret.\@ A linear function known to all trustees is chosen and is fixed in the beginning, which is used to reveal the secret from the linear variety.\@ The $n$ shadows are hyperplanes containing the liner variety. Moreover the hyperplanes are chosen to satisfy the condition that, the intersection of less than $t$ of them results in a linear variety which projects uniformly over the scalar field by the linear functional used for revealing the secret. The number $t$ is called the threshold. Thus as more shadows are known more information is revealed about the linear variety used to keep the secret, however no information is revealed until the threshold number of shadows are known. He had shown that Blakley's scheme and Karin's scheme are equivalent and provided algorithms to convert one scheme to another. He also stated that the schemes are all specialization of generalized linear threshold scheme. Brickell\cite{brickell1989some} also give a generalized notion of Shamir and Blackley's schemes using vector spaces.
   \vskip 2mm
Researchers have investigated $(t, n)$ threshold secret sharing extensively.\@ Threshold schemes that can handle more complex access structures have been described by Simmons \cite{simmons1992} like weighted threshold schemes, hierarchical scheme, compartmental secret sharing $etc$.\@ They were found a wide range of useful applications. Sreekumar {\it et al.,} \cite{sreekumar2009secret} in 2009, developed threshold schemes based on Visual cryptography.

\section{GENERALIZED SECRET SHARING }

\noindent In the previous section, we mentioned that any  $t$ of the $n$ participants should be able to determine the secret.\@ A more general situation is to specify exactly which subsets of participants should be able to determine the secret and which subset should not. In this section we give the secret sharing constructions based on generalized access structure. Shamir \cite{shamir1979} discussed the case of sharing a secret between the executives of a company such that the secret can be recovered by any three executives, or by any executive and any vice-president, or by the president alone.\@ This is an example of \textit{hierarchical secret sharing} scheme. The Shamir’s solution for this case is based on an ordinary $(3,m)$ threshold secret sharing scheme. Thus, the president receives three shares, each vice-president receives two shares and finally every  executive receives a single share.
   \vskip 2mm
The above idea leads to the so-called weighted(or multiple shares based) threshold secret sharing schemes. In these schemes, the shares are pairwise disjoint sets of shares provided by an ordinary threshold secret sharing scheme. Benaloh and Leichter have proven in \cite{benaloh1990generalized} that there are access structures that can not be realized using such scheme.
   \vskip 2mm
Several researchers address this problem and introduced secret sharing schemes realizing the general access structure.\@ The most effecient and easy to implement scheme was Ito, Saito, Nishizeki's \cite{ito1989secret} construction. It is based on Shamir's scheme. The idea is to  distribute shares to each authorized set of participants using multiple assignment scheme, where more than one share is assigned to a participant, if he belongs to more than one minimal authorized subset.
   \vskip 2mm
A simple scheme is mentioned by Beimel \cite{beimel2011secret}, in which the secret $S \in \{0,1\}$ and let $ \Gamma$ be any monotone access structure.\@ The dealer shares the secret independently for each authorized set
$ B \in \Gamma $, where $B=\{P_{i1},\ldots,P_{il}\}$.\@ The Dealer chooses $l-1$ random bits $r_{1},\ldots,r_{l-1}$.
Compute $r_{l}= S \oplus r_{1} \oplus r_{2} \oplus \cdots \oplus r_{l-1}$, and the Dealer distributes share $r_{j}$ to $P_{ij}$.\@ For each set $ B \in \Gamma$, the random bits are chosen independently and each set in $\Gamma$ can reconstruct the secret by computing the exclusive-or of the bits given to the set.\@ The unauthorized set cannot do so.
       \vskip 2mm
The disadvantage with multiple share assignment scheme is that the share size depends on the number of authorized set that contain $P_{j}$. A simple optimization is to share the secret $S$ only for minimal authorized sets. Still this scheme is inefficient for access structures in which the number of minimal set is big (Eg:$(n/2,n)$ scheme). The share size grows exponentially in this case.
        \vskip 2mm
Benalohand Leichter \cite{benaloh1990generalized} developed a secret sharing scheme for an access structure based on monotone formula.\@ This generalizes the multiple assignment scheme of Ito, Saito and Nishizeki \cite{ito1989secret}. The idea is to translate the monotone access structure into a monotone formula. Each variable in the formula is associated with a trustee in $\mathcal{P}$ and the value of the formula is \textit{true} if and only if the set of variables which are \textit{true} corresponds to a subset of $\mathcal{P}$ which is in the access structure. This formula is then used as a template to describe how a secret is to be divided into shares.
       \vskip 2mm
The monotone function contains only AND and OR operator. To divide a secret $S$ into shares such that  $P_{1} \; or \; P_{2}$ can reconstruct $S$. In this case $P_{1}$ and  $P_{2}$ can simply both be given values $S$. If $P_{1} \; and \; P_{2}$ need to reconstruct secret, then $P_{1}$ can be given value $S_{1}$ and $P_{2}$ can be given value $S_{2}$ such that $S=S_{1}+S_{2} \;mod \; m$,$(0 \le S \le m)$, $S_{1}$ is chosen randomly from $\mathbb{Z}_{m}$, $S_{2}$ is $(S-S_{1}) \; mod \; m$.
       \vskip 2mm
More exactly, for a monotone authorized access structure $\Gamma$ of size $n,$ they defined the set $\mathcal{F_A}$ as the set of formula on a set of variables $\{v_1,v_2,\ldots,v_n\}$ such that for every $\mathcal{F} \in \mathcal{F_A}$ the interpretation of $\mathcal{F}$ with respect to an assignation of the variables is true if and only if the true variables correspond to a set $A \in \Gamma$. They have remarked that such formula can be used as templates for describing how a secret can be shared with respect to the given access structure. Because the formula can be expressed using only `$\wedge$' operators and `$\vee$' operators, it is sufficient to indicate how to ``split'' the secret across these operators.
   \vskip 2mm
Brickell \cite{brickell1991classification} developed some ideal schemes for generalized access structure using vector spaces. Stinson \cite{stinson1992explication} introduced a monotone circuit construction based on monotone formula and also the construction based on public distribution rules. Benaloh's scheme was generalized by Karchmer and Wigderson \cite{karchmer1993span}, who showed that if an access structure can be described by a small monotone span program then it has an efficient scheme.
         \vskip 2mm
Cumulative schemes were first introduced by Ito {\it et al.,} \cite{ito1989secret} and then used by several authors to construct a general scheme for arbitrary access structures.\@ Simmons \cite{simmons1992} proposed cumulative map, Jackson \cite{jackson1993cumulative} proposed a notion of cumulative array. Ghodosi {\it et al.,} \cite{ghodosi1998construction} introduced simpler and more efficient scheme and also introduced capabilities to detect cheaters. Generalized cumulative arrays in secret sharing is introduced by Long \cite{long2006generalised}.
 
\section{MULTI SECRET SHARING}
\noindent There are several situations in which more than one secret is to be shared among participants. As an example, consider the following situation, described by Simmon \cite{simmons1992}.\@ There is a missile battery and not all of the missiles have the
same launch enable code.\@ We  have to devise a scheme which will allow  any selected subset of users to enable different launch code.\@ The problem is to devise a scheme which will allow any one, or any selected subset, of the launch enable codes to be activated in this scheme.\@ This problem could be trivially solved by realizing different secret sharing schemes, one for each of the launch enable codes, but this solution is clearly unacceptable since each participant should remember too much information. What is really needed is an algorithm such that the same pieces of private information could be used to recover different secrets. 
   \vskip 2mm
One common drawback of all secret  sharing scheme is that, they are one-time schemes. That is once a qualified group of participants reconstructs the secret $K$ by pooling their shares, both the secret $K$ and all the shares become known to everyone, and there is no further secret. In other words, the share kept by each participant can be used to reconstruct only one secret.
   \vskip 2mm
Karnin, Greene and Hellman \cite{karnin1983} in 1983 mentioned the  multiple secret sharing scheme where threshold number of users can reconstruct multiple secrets at the same time. Alternatively the scheme can be used to share a large secret by splitting it into smaller shares. Franklin {\it et al.,} \cite{franklin1992communication}, in 1992 used a technique in which the polynomial-based single secret sharing is replaced with a scheme where multiple secrets are kept hidden in a single polynomial.\@ They also considered the case of dependent secrets in which the amount of information distributed to any participant is less than the information distributed with independent schemes. Both the schemes are not perfect. They are also one time threshold schemes. \@ That is, the shares cannot be reused.
   \vskip 2mm
Blundo {\it et al.,} \cite{blundo1993efficient}, in 1993 considered the case in which $m$ secrets are shared among participants in a single access structure $\Gamma$ in such a way that any qualified set of participants can reconstruct the secret.\@ But any unqualified set of participants knowing the value of number of secrets might determine some (possibly no) information on other secrets. Jackson {\it et al.,} \cite{jackson1994multisecret}, in 1994  considered the situation in which there is a secret $S_k$ associated with each subset $k$ of participants and $S_k$ can be reconstructed by any group of $t$ participants in $k$ $(t\le k)$.\@ That is each subset of $k$ participants is associated with a secret which is protected by a $( t , k$)-threshold access structure.\@ These schemes are called  multi-secret threshold schemes.\@ They came up with a combinatorial model and optimum threshold multi secret sharing scheme. Information theoretic model similar to threshold scheme is also proposed for multi-secret sharing.\@ They have generalized and classified the multi-secret sharing scheme based on the following facts.
\begin{itemize}
\item{Should all the secrets be available for potential reconstruction during the lifetime of the scheme, or should the access of secrets be further controlled by enabling the  reconstruction of a particular secret only after extra information has been broadcast to the participants.}
\item{Whether the scheme can be used just once to enable the secrets or should the scheme be designed to enable multiple use. }
\item{If the scheme is used more than once then the reconstructed secret or shares of the participants is known to all other participants or it is known to only the authorized set.} 
\item{The access structure is  generalized or threshold in nature.}
\end{itemize}

In 1994 He and Dawson \cite{he1995multisecret} proposed the general implementation of  multistage secret sharing. The proposed scheme allows many secrets to be shared in such a way that all secrets can be reconstructed separately. The implementation uses Shamir's threshold scheme and assumes the existence of a one way function which is hard to invert.\@ The public shift technique is used here. A $t-1$ degree polynomial $f(x)$ is constructed first, as in Shamir's scheme.\@ The public shift values are $d_i=z_i-y_i$, where $z_i=f(x_i)$. The $y_i$'s are the secret shares of the participant. $y_i$'s are then send to the participants secretly. For sharing the next secret, $h(y_i)$ is used, where $h$ is the one way function. The secrets are reconstructed in particular order, stage by stage and also this scheme needs $kn$ public values corresponds to the $k$ secrets. The advantage is that each participant has to keep only one secret element and is of the same size as any shared secret.\@ In 1995 Harn \cite{harn1995efficient} shows an alternative implementation of multi stage secret sharing which requires only $k(n-t)$ public values. The implementation become very attractive, especially when the threshold value $t$ is very close to the number of participants $n$. In this scheme an $(n-1)$ degree polynomial $f(x)$ is evaluated at $(n-t)$ points and are made public. Any $t$ participants can combine their shares with the $(n-t)$ public shares to interpolate the degree $(n-1)$ polynomial. Multiple secrets are shared with the help of one way function as in He and Dawson scheme.
   \vskip 2mm
The desirable properties of a particular scheme depends on both the requirements of the application and also the implementation. Several multi secret threshold schemes are developed by the research community. In this survey we only explore some of the important constructions of multi-secret sharing scheme realizing general access structure.

\subsection{Cachin's Scheme}

\noindent A computationally secure secret sharing scheme with general access structure, where all shares are as short as the secret is proposed by Christian Cachin \cite{cachin1995line} in 1995.\@ The scheme also provides capability to share multiple secrets and to dynamically add participants on-line without having to redistribute new shares secretly to the current participants. These capabilities are achieved by storing additional authentic information in a publicly accessible place which is called a noticeboard or bulletin board. This information can be broadcast to the participants over a public channel. The protocol gains its security from any one-way function.The construction has the following properties.

\begin{itemize}
\item{All shares must be transmitted and stored secretly once for every participants and are as short as the secret.}
\item{Multiple secret can be shared with different access structure requiring only one share per participant for all secrets.}
\item{Provides the ability for the dealer to change the secret after the shares have been distributed.}
\item{The dealer can distribute the shares on-line. When a new participant is added and the access structure is changed, already distributed shares remain valid. Shares must be secretly send to the new participants and the publicly readable information has to be changed.}
\end{itemize} 
Let the secret $K$ be an element of finite Abelian Group $\textbf{G}=<G,+>$. The basic protocol to share a single secret is as follows.
\begin{enumerate}
\item{The dealer randomly chooses $n$ elements $S_1,S_2,\ldots,S_n$ from $G$ according to the uniform distribution and send them secretly to the participants over a secret channel.}
\item{For each minimal qualified subset $X \in \varGamma_0$}, the dealer computes $$T_X=K-f(\sum_{x:P_{x} \in X}S_X)$$
and publishes $\mathcal{T}={T_X|X \in  \varGamma_0}$ on the bulletin board.
\end{enumerate}
In order to recover the secret $K$, a qualified set of participants $Y$ proceeds as follows.
\begin{enumerate}
\item{The members of $Y$ agree on a minimal qualified subset $X {\subseteq} Y$.}
\item{The members of $X$ add their shares together to get $V_X=\sum_{x:P_x \in X}{S_X}$ and apply the one-way function $f$ to the result.}
\item{They fetch $T_X$ from the bulletin board and compute $K=T_X+f(V_X)$}
\end{enumerate}

The shares of the participants in $X$ are used in the computation to recover the secret $K$. For the basic scheme where only one secret is shared, the shares do not have to be kept secret during this computation. However for sharing multiple secrets the shares and the result of their addition have to be kept secret. \\ \\
In order to share multiple secrets $K^1,K^2,\ldots,K^h$ with different access structures $\Gamma^1,\Gamma^2,\ldots,\Gamma^h$ among the same set of participants $\mathcal{P}$, the dealer has to distribute the private shares $S_i$ only once but prepares $\Gamma^1,\Gamma^2,\ldots,\Gamma^h$ for each secret. The single secret sharing scheme cannot be applied directly for multi secret sharing because it is not secure. If a group of participants $X$ qualified to recover both $K^1$ and $K^2$ then any group $Y \in \Gamma^1$ can obtain $K^2$ as $$K^2=T_X^{2}+T_Y^{1}+f(V_Y)-T_X^{1}$$
   \vskip 2mm
To remedy this deficiency, the function $f$ is replaced by a family $F={f_h}$ of one-way functions so that different one-way functions are employed for different secrets. The following protocol is used to share $m$ secrets.
\begin{enumerate}
\item{The dealer randomly chooses $n$ elements $S_1,S_2,\ldots,S_n$ from $G$ and send them securely to the participants as shares.}
\item{For each secret $K^h$ to share( with $h=1,\ldots,m$) and for each minimal qualified subset $X \in \Gamma_0^h$, the dealer computes $$T_X^h=K^h-f_h(\sum_{x:P_x\in X}S_x)$$ and publishes $\mathcal{T}^h=\{T_X^h|X \in \Gamma_0^h\}$ on the bulletin board.}
\end{enumerate}
In order ro recover some secret $K^h$, a set of participants $Y \in \varGamma^h$ proceeds as follows.
\begin{enumerate}
\item{The members of $Y$ agree on a minimal qualified subset $X {\subseteq} Y$.}
\item{The members of $X$ add their shares together to get $V_X=\sum_{x:P_x \in X}{S_X}$ and apply the one-way function $f_h$ to the result.}
\item{They fetch $T_X^h$ from the bulletin board and compute $K^h=T_X^h+f_h(V_X)$}
\end{enumerate}
The scheme does not demand a particular order for the reconstruction of the secrets as in He and Dawson scheme.\@ The required family of functions $F$ can be easily be obtained from $f$ by setting $f_h(x)=f(h+x)$, when $h$ is represented suitably in $G$. Because  different one-way function $f_h$ is used for each secret, it is computationally secure. But the shares have to be protected from the eyes of other participants during the reconstruction.\@ Otherwise, these participants could subsequently recover other secrets they are not allowed to know. Therefore the computation of $f_h(V_X)$ should be done with out revealing the secret shares.
   \vskip 2mm
In many situations, the participant of a secret sharing scheme do not remain the same during the entire life-time of the secret. The access structure may also change. In this scheme it is assumed that the changes to the access structure are monotone, that is participants are only added and qualified subsets remain qualified.\@ The scheme is not suitable for access structures which are non-monotonic. Removing participants is also an issue which is not addressed. In multi-secret sharing, the shares must be kept hidden to carry out the computation. Cachin suggest that computations involved in recovering $K$ could be hidden from the participants, using a distributed evaluation protocol proposed by Goldreich {\it et al.,} \cite{goldreich1987play}. For access to a predetermined number of secrets in fixed order, a variant of one-time user authentication protocol of Lamport \cite{lamport1981password}could be used. 

The proposed scheme has many practical applications in situations where the participants and the access rules or the secret itself frequently change. No new shares have to be distributed secretly when new participants are included or participants leave. Such situation often arise in key management, escrowed system etc.

\subsection{Pinch's Scheme}
\noindent The Cachin's scheme does not allow shares to be reused after the secret has been reconstructed. A distributed computation sub protocol is proposed using one way function but it allows the secret to be reconstructed in a specified order. Pinch  \cite{pinch1996line} in 1996 proposed a modified algorithm based on the intractability of the Diffie-Hellman problem, in which arbitrary number of secrets can be reconstructed without having to redistribute new shares.
   \vskip 2mm
Let $M$ be a multiplicative group in which the Diffie-Hellman problem is intractable.\@ That is, given elements $g,\;g^x\;\mbox{and}\; g^y$ in $M$ it is computationally infeasible to obtain $g^{xy}$.\@ This implies the intractability of the discrete logarithm problem. If the discrete logarithm problem can be solved then the Diffie-Hellman problem can also be solved. Suppose $f:M\implies G$ is a one-way function, where $G$ be the additive group modulo some prime $p$ and $M$ be the multiplicative group to the same modulus, which will be cyclic of order $q$. The protocol proceeds as follows:
\begin{enumerate}
\item {The dealer randomly chooses secret shares $S_i$,  as integers coprime to $q$, for each participant $P_i$ and send them through a secure channel. Alternatively Diffie-Hellman key exchange can be used using the group $M$ to securely exchange $S_i$.}
\item {For each minimal trusted set $X \in \Gamma$, the dealer randomly chooses $g_X$ to be a generator of $M$ and computes $$T_X=K-f\left(g_X^{\prod_{x \in X}S_x}\right)$$} and publish $(g_X,T_X)$ on the notice board.
\end{enumerate}
In order to recover the secret $K$, a minimal trusted set $X={P_1,\ldots,P_t}$, of participants comes together and follow
the protocol mentioned below.
\begin{enumerate}
\item{Member $P_1$ reads $g_X$ from the notice board and computes $g_X^{S_1}$ and passes the result to $P_2$.}
\item{Each subsequent member $P_i$, for $1<i<t$, receives $g_X^{S_1\cdots S_{i-1}}$ and raises this value to the power $S_i$ to form $$V_X=g_X^{\prod_{i=1}^{t}S_i}=g_X^{\prod_{x \in X}S_x}$$}
\item {On behalf of the group $X$, the member $P_t$ reads $T_X$ from the notice board and can now reconstruct $K$ as $K=T_X+f(V_X)$.}
\end{enumerate}
If there are multiple secrets $K_i$ to share, it is now possible to use the same one way function $f$, provided that each entry on the notice board has a fresh value of $g$ attached.\@ There is a variant proposal which avoids the necessity for the first participant to reveal $g^{S_1}$ at the first step.\@ The participant $P_1$ generates a random $r$ modulo $q$ and passes the result of  $g^{rS_1}$ to $P_2$. The participant $P_t$ will pass $g_X^{rS_1 \cdots S_{t}}$ back to $P_1$. $P_1$ can find $w$ such that $rw \equiv 1 \; \mbox{mod}\; q$ and raises $g_X^{rS_1 \cdots S_n}$ to the power $w$ to form $$V_X=g_X^{\prod_{i=1}^{t}S_i}=g_X^{\prod_{x \in X}S_x}$$
   \vskip 2mm
Ghodosi {\it et al.,} \cite{ghodosi1997prevent} showed that Pinch's scheme is vulnerable to cheating and they modified the scheme to include cheating prevention technique. In Pinch's scheme a dishonest participant $P_i \in X$ may contribute a fake share $S_i^{'}=\alpha S_i$, where $\alpha$ is a random integer modulo $q$. Since every participant of an authorized set has access to the final result $g_X^{S_1,\cdots,S_i^\prime,\cdots,S_t}$, the participant $P_i$ can calculate the value $${\left(g_X^{S_1,\cdots,S_i^\prime,\cdots,S_t}\right)}^{\alpha^{-1}}=$$ \\ $$g_X^{S_1,\cdots,S_i,\cdots, S_t}=g_X^{\prod_{x \in X} S_x}=V_X$$
and hence obtain the correct secret, where as the other participants will get an invalid secret.
   \vskip 2mm
The cheating can be detected by publishing $g_X^{V_X}$ corresponds to the every authorized set $X$ in the initialization step by the dealer. Every participants $x \in X$ can verify whether $g_X^{V_X} =g_X^{V_X^ \prime}$, where $V_X^\prime$ is the reconstructed value.\@ However this cannot prevent cheating or cheaters can be identified.\@ The cheating can be prevented by publishing extra information on the notice board.\@ Let $C= \sum_{x \in X}g_x^{S_x}$.\@ For each authorized set $X$, the dealer also publishes $C_X=g_X^C$.\@ At the reconstruction phase, every participant $P_i \in X$ computes $g_x^{S_i}$ and broadcasts it to all participants in the set $X$. Thus every participant can computes $C$ and verifies $C_X=g_X^C$. If the verification fails, then the protocol stops.
If there exist a group of collaborating cheats, they can cheat in the  first stage. Yeun {\it et al.,} \cite{yeun1998identify} proposed a modified version of the Pinch's protocol which identifies all cheaters regardless of their number, improving on previous results by Pinch and Ghodosi {\it et al.}
\subsection{RJH and CCH scheme}
\noindent An efficient computationally secure on-line secret sharing scheme is proposed by Re-Junn Hwang and Chin-Chen Chang  \cite{hwang1998line} in 1998.\@ In this each participant hold a single secret which is as short as the shared secret.\@ They are selected by the participants itself, so a secure channel is not required between the dealer and the participants. Participants can be added or deleted and secrets can be renewed with out modifying the secret share of the participants. The shares of the participants is kept hidden and hence can be used to recover multi secrets. The scheme is multi use unlike the one time multi secret sharing scheme.
   \vskip 2mm
In Cachin's and Pinch's schemes, the dealer has to store the shadow of each participant to maintain the on-line property. The dealer storing the shares is an undesirable property in secret sharing scheme. This scheme avoids the problem and provides great capabilities for many applications.\@ The scheme has four phases:initialization phase, construction phase, recovery phase and reconstruction/renew phase.
   \vskip 2mm
Assume that there are $n$ participants $P_1,P_2,\ldots,P_n$, sharing a secret $K$ with the monotone access structure $\Gamma=\{\gamma_1,\gamma_2,\ldots,\gamma_t\}$. In the initialization phase the dealer select two strong primes $p$ and $q$  and publishes $N$ on the public bulletin,  where $N$ is the multiplication of $p$ and $q$.\@
The dealer also chooses another integer $g$ from the interval $[N^{1/2},N]$ and another prime $Q$ which is larger than $N$ and publishes them. Each participant can select an integer $S_i$ in the interval $[2,N]$ and computes $U_i=g^{S_i}\; \mbox{mod}\; N$.\@ Each participant keeps $S_i$ secret and send the pseudo share $U_i$ and the identifier $ID_i$ to the dealer.\@ If certain different participant select same shadow, the dealer asks for new shadows or alternatively the dealer can select the shares and send to the participants securely.\@ But this need a secure channel. Finally  dealer publishes $(ID_i,U_i)$ of each participant $P_i$ in the public bulletin.
   \vskip 2mm
In the construction phase the dealer computes and publishes some information for each qualified subset in access structure $\Gamma$. The participants of any qualified subset $\gamma_j$ can cooperate to recover the shared secret $K$ by using these information and the values generated from their shadows in the recovery phase. The public information corresponds to each qualified set is generated as follows.
\begin{itemize}
\item Randomly select an integer $S_0$ from the interval $[2,N]$ such that $S_0$ is relatively prime to $p-1$ and $q-1$.
\item Compute $U_0=g^{S_0} \mbox{mod}\;N$ and $U_0 \neq U_i$ for all $i=1,2,\ldots,n.$
\item Generate an integer $h$ such that $S_0\times h=1\;\mbox{mod}\; \phi(N).$
\item Publish $U_0$ and $h$ on the public bulletin.
\item For each minimal qualified subset $\gamma_j=P_{j1},P_{j2},\ldots,P_{jd}$ of $\Gamma_0$,  the dealer computes public information $T_j$ as follows.
\item Compute $H_j=K \oplus (U_{j1}^{S_0} \;\mbox{mod}\; N) \oplus (U_{j2}^{S_0} \;\mbox{mod}\; N)\; \oplus, \ldots ,\oplus (U_{jd}^{S_0} \;\mbox{mod}\; N)$
\item Use $d+1$ points $(0,H_j),(ID_{j1},(U_{j1}^{S_0}\; \mbox{mod}\; N)),\\ \ldots,(ID_{jd},(U_{jd}^{S_0}\; \mbox{mod}\; N))$ to construct a polynomial $f(X)$ of degree $d$
\begin{equation*}
\begin{split}
 f(x)&=H_j \times \prod_{k=1}^{d}(X-ID_{jk})/(-ID_{jk})+  \\
 & \sum_{l=1}^{d}[(P_{jl}^{S_0}\;\mbox{mod}\; N)  \times (X/ID_{jl})\times \\ &\prod_{\substack{k=1\\k \ne l}}^{d}(X-ID_{jk})/(ID_{jl}-ID_{jk})] \;\mbox{mod}\; Q
 \end{split}
 \end{equation*}
 where $d$ is the number of participants in qualified subset $\gamma_j$
 \item Compute and publish $T_j=f(1)$ on the public bulletin.
\end{itemize}

In the recovery phase participants of any qualified subset can cooperate to recover the shared secret $K$ as follows.
\begin{itemize}
\item Each participant gets $(U_0,h,N)$ from the public bulletin.
\item Each participant $P_{ij}$, computes and provides 
${S_{ji}}^{'}=U_0^{S_{ji}}\;\mbox{mod}\;N$ ,where ${S_{ji}}^{'}$ is the pseudo share of $P_{ji}$.
$S_{ji}^{'h} \;\mbox{mod}\; N=U_{ji}$, then $S_{ji}^{'}$ is the true shadow else it is false and the participant $P_{ji}$ is the cheater.
\item Get $T_j$ from the public bulletin and use $d+1$ points $(1,T_j),(ID_{j1},S_{j1}^{'}),\ldots,(ID_{jd},S_{jd}^{'})$ and use Lagrange interpolation to reconstruct the $d$ degree polynomial $f(X)$:
\begin{equation*}
\begin{split}
 f(X)&=T_j \times \prod_{k=1}^{d}(X-ID_{jk})/(1-ID_{jk})+  \\
 & \sum_{l=1}^{d}[(S_{jl}^{'} \times (X-1/ID_{jl}-1)\times \\ &\prod_{\substack{k=1\\k \ne l}}^{d}(X-ID_{jk})/(ID_{jl}-ID_{jk})] \;\mbox{mod}\; Q
 \end{split}
 \end{equation*}
\item Compute $H_j=f(0)$ and recover the secret $K=H_j \oplus S_{j1}^{'} \oplus S_{j2}^{'} \oplus \cdots 
\oplus S_{jd}^{'}$
\end{itemize}

When new participants join the group, the access structure changes.\@ The dealer then performs the construction phase and publish the new public information.\@ The older participants share remain the same.\@ When the participants disenrolled, the corresponding minimal qualified subset should be deleted from the access structure. The shared secret should be renewed for security consideration.\@ Public information must be changed in this case but the rest of the authorized participants still hold the same shadows. Changing the shared secret can also be done by modifying the public values but the same shadows can be reused.
   \vskip 2mm
Adding a new subset can also be done easily. If the new qualified subset contains an old minimal qualified subset in the access structure, then nothing needs to be done. If there are old minimal qualified subsets in the new qualified subset, the old ones shall be deleted from the access structure and the public information is updated according to the new access structure.\@ Canceling a qualified subset needs the shared secret to be renewed. The public information corresponds to the rest of the qualified subset must be modified.\@ The public information corresponds to the canceled subset is of no use and is removed.\@ It is noted that the dealer does not need to collect the shadows of all the participants to reconstruct the secret sharing scheme again.
   \vskip 2mm
To share multiple secrets $K_1,K_2,\ldots,K_n$ with the access structure $\Gamma_1,\Gamma_2,\ldots,\Gamma_n$,    each participant holds only one share $S_i$ for these $n$ secrets.\@ For each shared secret $K_i$ the dealer select a unique $S_0^{i}$ and publishes the corresponding ${h_i, U_{0i}}$.\@ The dealer also generate and publishes the information $T_{ij}$ for each qualified subset $\gamma_{ij}$ in minimal access structure $\Gamma_i$. The participants of each qualified subset $\gamma_{ij}$ in $\Gamma_i$ can cooperate to recover the shared secret $K_i$ by performing the recovery phase.
\subsection{Sun's Scheme}
\noindent In Pinch's scheme high computation overhead is involved and also sequential reconstruction is used in the recovery phase. In 1999 Sun \cite{sun1999line} proposed a scheme having the advantages of lower computation overhead and parallel reconstruction in the secret recovery phase.\@ The security of the scheme is only based on one-way function, not on any other intractable problem.
   \vskip 2mm
Let $f$ be a one way function with both domain and range $G$. The following protocol is used to share $m$ secrets $K^{[h]}$ with access structures $\Gamma^{[h]}$ for $h=1,\ldots,m$.
\begin{enumerate}
\item{The dealer randomly chooses $n$ secret shares $S_i,\ldots,S_n$ and send them to the participants through a secret channel.}
\item{For every shared secret $K^{[h]}$ and for every minimal qualified subset $X \in \Gamma_0^{[h]}$, the dealer randomly chooses $R_X^{[h]}$ in $G$ and computes $$ T_X^{[h]}=K^{[h]} - \sum_{x:P_x \in X}f(R_X^{[h]} + S_x)$$ and publishes $H^{[h]}=\{(R_X^{[h]},T_X^{[h]})|X \in \Gamma_0^{[h]}\}$ on the notice board.}
\end{enumerate}
In order to recover the secret $K^{[h]}$, a set of participants $Y \in \Gamma^{[h]}$ proceeds as follows
\begin{enumerate}
\item{The members of $Y$ agree on a minimal qualified subset $X \subseteq Y$, where $X=\{P1,\ldots,P_t\}$}

\item{Each member $P_i$ reads $R_X^{[h]}$ from the notice board and computes $f(R_X^{[h]}+ S_i)$ and send the result to $P_t$ who is designated as secret re-constructor.}
\item{$P_t$ receives $f(R_X^{[h]}+ S_i)$ for $1 \le i \le t-1$, and reconstructs the secret $K^{[h]}=T_X^{[h]}+ \sum_{i=1}^{t}f(R_X^{[h]}+S_i)$}
\end{enumerate}
Once the secret is reconstructed it become public.\@ $f(R_X^{[h]}+ S_i)$ is unique for every secret and every authorized set.\@ Most of the implementations of one way functions are based on permutations, substitution and XOR operation.\@ Therefore the computation is much faster than the exponentiation.\@ The step2 of the reconstruction phase can proceed parallelly where as in Pinch's scheme the construction is sequential.\@ Cheating can be detected by putting additional information $f(K^{[h]})$ on the notice board for every shared secret.\@ Any one can verify the correctness of the computed secret.\@ The scheme can also detect cheaters by putting additional information $C_{X,i}^{[h]}=f(f(R_X^{[h]}+S_i))$ for  every secret $K^{h}$, every authorized set $X$ and for every participant $P_i$. The scheme is dynamic. Participants or new access structure can be added by distributing shares to the new participants and update public information on the notice board. The previously distributed shares remain valid. When some participants
or some access structures need to be deleted, the shared secret should be renewed. The dealer only need to update the information on bulletin board.	
\begin{table*}[!htb]
\renewcommand{\baselinestretch}{1}
\caption{Comparison of Multi secret sharing schemes \label{tab:comp}}
\begin{small}
\begin{center}
\begin{tabular}{|p{3.5cm}|c|c|c|c|c|c|} \hline
  Properties  & Cachin \cite{cachin1995line} & Pinch \cite{pinch1996line} & RJH  CCH \cite{hwang1998line} & Sun \cite{sun1999line} & Das \cite{das2010efficient} & Roy \cite{roy2010multi}  \\ \hline
  share size same as secret & Yes  	& Yes & Yes & Yes  & Yes & Yes  \\ \hline
  use of one way function	& Yes   & Yes &  No & Yes  & Yes & Yes  \\ \hline
  use of discrete logarithm	& No    & Yes &  Yes & No  & No  & No  \\ \hline
  use of interpolation 		& No    & No  &  Yes & No  & No  & Yes  \\ \hline
  shares remain secret during reconstruction & No    & Yes  &  Yes & Yes  & Yes  & Yes  \\ \hline
  dealer knows the share    & Yes   & Yes  &  No & Yes  & Yes  & Yes  \\ \hline
  shares can be reused		& No   & Yes  &  Yes & Yes  & Yes  & Yes  \\ \hline
  dynamic					& No   & Yes  &  Yes & Yes  & Yes  & Yes  \\ \hline
  verifiability				& No   & No  &  Yes & Yes  & Yes  & Yes  \\ \hline
\end{tabular}
\end{center}
\end{small}
\end{table*}
\subsection{Adhikari {\it et al.,} Scheme}
\noindent An efficient, renewable, multi use, multi-secret sharing scheme for general access structure is proposed by Angsuman Das and Avishek Adhikari  \cite{das2010efficient} in 2010.\@ The scheme is based on one way hash function and is computationally more efficient.\@ Both the combiner and the participants can also verify the correctness of the information exchanged among themselves in this.\@ The scheme consist of three phases. The dealer phase, pseudo-share generation phase and the combiner's phase.
   \vskip 2mm
Let $\mathcal{P}=\{P_1,P_2,\ldots,P_n\}$ be the set of participants and $S_1,S_2,\ldots,S_k$ be the $k$ secrets to be shared by a trusted dealer. Each secret is of size $q$ bits.\@ $\Gamma_{S_i}=\{A_{i1},A_{i2},\ldots,A_{it}\}$ be the access structure corresponds to the secret $S_i$ and $A_{il}$ is the $l$'th qualified subset of the access structure of the $i$'th secret $S_i$
   \vskip 2mm
In the dealer phase, the dealer $\mathcal{D}$ chooses a collision resistant one-way hash function $H$, which takes as argument a binary string of arbitrary length and produces an output a binary string of fixed length $q$, where $q$ is the length of each secret. The dealer also choose randomly $x_\alpha$ the shares of size $q$ and send to the participants through a secure channel.
   \vskip 2mm
In the pseudo share generation phase, a pseudo share corresponds to each secret and for each authorized set is generated from the participants secret share in the following way $$S_{ij} = S_i \bigoplus \left\{\bigoplus_{\alpha:P_\alpha \in A_{ij}} H(x_\alpha \parallel i_l \parallel j_m) \right \}$$
where $i_l$ represent the $l$ bit representation of the number of secret ie; $l= \lfloor log_2k \rfloor + 1$ and $m= \lfloor log_2t \rfloor+1$ , $t$ is the maximum size of an authorized subset among the access structures corresponds to different secrets.The dealer then publishes the values $S_{ij},H(S_i),H^2(x_\alpha \parallel i_l \parallel j_m)$
   \vskip 2mm
In the combiners phase the participants of an authorized subset $A_{ij}$ of $\Gamma_{S_i}$ submit the pseudo share $H(x_\alpha \parallel i_l \parallel j_m)$ which is then x-or with $S_{ij}$ to get the secret $S_i$ by the combiner.$$S_{i} = S_{ij} \bigoplus \left\{\bigoplus_{\alpha:P_\alpha \in A_{ij}} H(x_\alpha \parallel i_l \parallel j_m) \right \}$$The combiner can verify the pseudo share given by the participant by checking it with the public value $H^2(x_\alpha \parallel i_l \parallel j_m)$. The participants can check whether the combiner is giving them back the correct secret $S_i$ by verifying it with the public value $H(S_i)$.
   \vskip 2mm
Adhikari and Roy \cite{roy2010multi} also proposed a similar scheme with polynomial interpolation. In this scheme, for each authorized subset in the access structure corresponds to a secret, a polynomial of degree $m-1$ is created with the constant term as the secret $S_i$, where $m$ is the number of participants in the authorized subset.
$$f_q^{S_{i}}(x)= S_i+d_1^{i_q}x+d_2^{i_q}x^2+
\ldots+d_{mi_q-1}^{i_q}x^{m_{i_q}-1}$$
For each participant $P_b^{i_q} \in A_q^{S_i}$ in $\Gamma_{S_i} $ the dealer compute pseudo share $U_{P_b}^{i_q}=h(x_{P_b^{i_q}})\parallel i_l \parallel q_m$, where $x_i$ is the secret share of the participant and 
$i=1,\ldots,k; q=1,\ldots,l; b=1,\ldots,m$. The dealer also computes $B_{P_b}^{i_q}=f_q^{S_i}(ID_b^{i_q})$. Finally the shift values are computed and published corresponds to  each secret and each authorized subset $M_{P_b}^{i_q}=B_{P_b}^{i_q}-U_{P_b}^{i_q}$.
   \vskip 2mm
In the reconstruction phase the pseudo shares of authorized set of participant can be added with the public information to obtain $B_{P_b}^{i_q}=f_q^{S_i}(ID_b^{i_q})=M_{P_b}^{i_q}+U_{P_b}^{i_q}$. The secret can be reconstructed by interpolation using these $m$ values.\\
$S_i=\sum_{b \in \{1,2,\ldots, m_{i_q}\}} B_{P_b}^{i_q}\prod_{r \in \{1,2,\ldots,m_{i_q}r \ne b\}}$ \\$\frac{-ID_{P_{r}^{i_q}}}{ID_{P_{b}^{i_q}}-ID_{P_{r}^{i_q}}},$
It is noted that the computational complexity is more in this case, compared with the previous scheme.

\section{SUMMARY}
In this section we give a brief summary of the important constructions for multi-secret sharing corresponds to generalized access structures. The table \ref{tab:comp} summarize and compares the important properties of different schemes.\@ The  important technique used for the constructions are based on one way functions, discrete logarithm problem and Shamir's secret sharing technique. The schemes based on discrete logarithm problem and hash functions provide only computational security because the security depends on the computational complexity of these problems. But for many of the cryptographic application with polynomial time bounded adversary, the computational security is sufficient.\@ For maintaining the unconditional security, large number of shares must be kept by the participant. The number of shares that must be kept is proportional to the number of secret to be shared.
   \vskip 2mm
The public values in the bulletin board of each scheme is proportional to the number of authorized subset in an access structure corresponds to each key.\@ There will be at least one public value corresponds to each authorized subset in the access structure corresponds to a key.\@ There are also additional public parameters used for the security of the scheme.\@ The computational complexity depends on the complexity of the one way function used or  the modular exponentiation.\@ But these operations can be efficiently done in polynomial time. The most commonly used one way functions like LFSR, MD5, SHA are all based on simple xor, permutation and substitution operation.\@ So these schemes can be implemented in polynomial time. Modular exponentiation is time consuming with large exponent but efficient algorithm exist for the fast computation. The share generation and reconstruction in the Shamir's scheme, which uses polynomial interpolation can also be implemented efficiently.
   \vskip 2mm
All the scheme mentioned assumes that the dealer is a trusted person. Cheating detection mechanisms are also proposed in some schemes with the help of additional public parameters. The combiner can verify the share submitted by the participants and the participant can also check the reconstructed secret. However the security is computational. If the computational problem is solved, the secret can be revealed by an adversary.The mathematical model, security notions and computational security for multi-secret sharing is proposed by Javier Herranz {\it et al.,}  \cite{herranz2013new}~\cite{herranz2013sharing} in 2013.
\section{CONCLUSIONS}
We have explored some important multi-secret sharing techniques for generalized monotone access structure in this survey.\@ There are several threshold multi-secret sharing schemes where multiple secrets are shared, each with different threshold.\@ These schemes are not considered here.\@ The emphasis is given to a more generalized notion, where each secret is shared according to a monotone generalized access structure.\@ Threshold multi-secret sharing also found several applications and we prefer users to further look into it.\@ The major concern in the multi-secret sharing is the large number of public values and the computational complexity.\@ Only computational security can be achieved in all the schemes mentioned, where security depends on the security of some computationally hard problem.\@ Multi-secret sharing schemes have found numerous application in implementing authentication mechanisms, resource management in cloud, multi policy distributed signatures, multi policy distributed decryption $etc$..
\small\balance

\small\balance
\noindent{\includegraphics[width=1in,height=1.7in,clip,keepaspectratio]{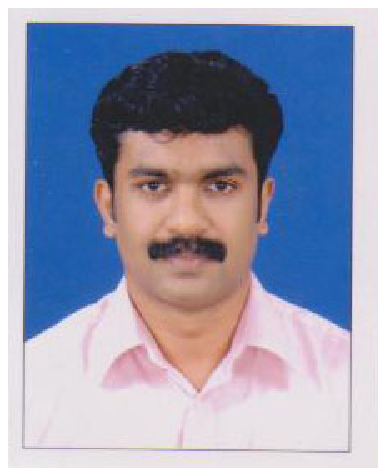}}
\begin{minipage}[b][1in][c]{1.8in}
{\centering{\bf {V P Binu}} is a Research Scholar in the Department of Computer Applications, Cochin University of Science and Technology(CUSAT). He Holds a Bachelor Degree in Computer Science and Engineering and Masters Degree  in Computer } \\
\end{minipage}
and Information Science. His research area includes Cryptography, Secret Sharing and Security. 

\noindent{\includegraphics[width=1in,height=1.7in,clip,keepaspectratio]{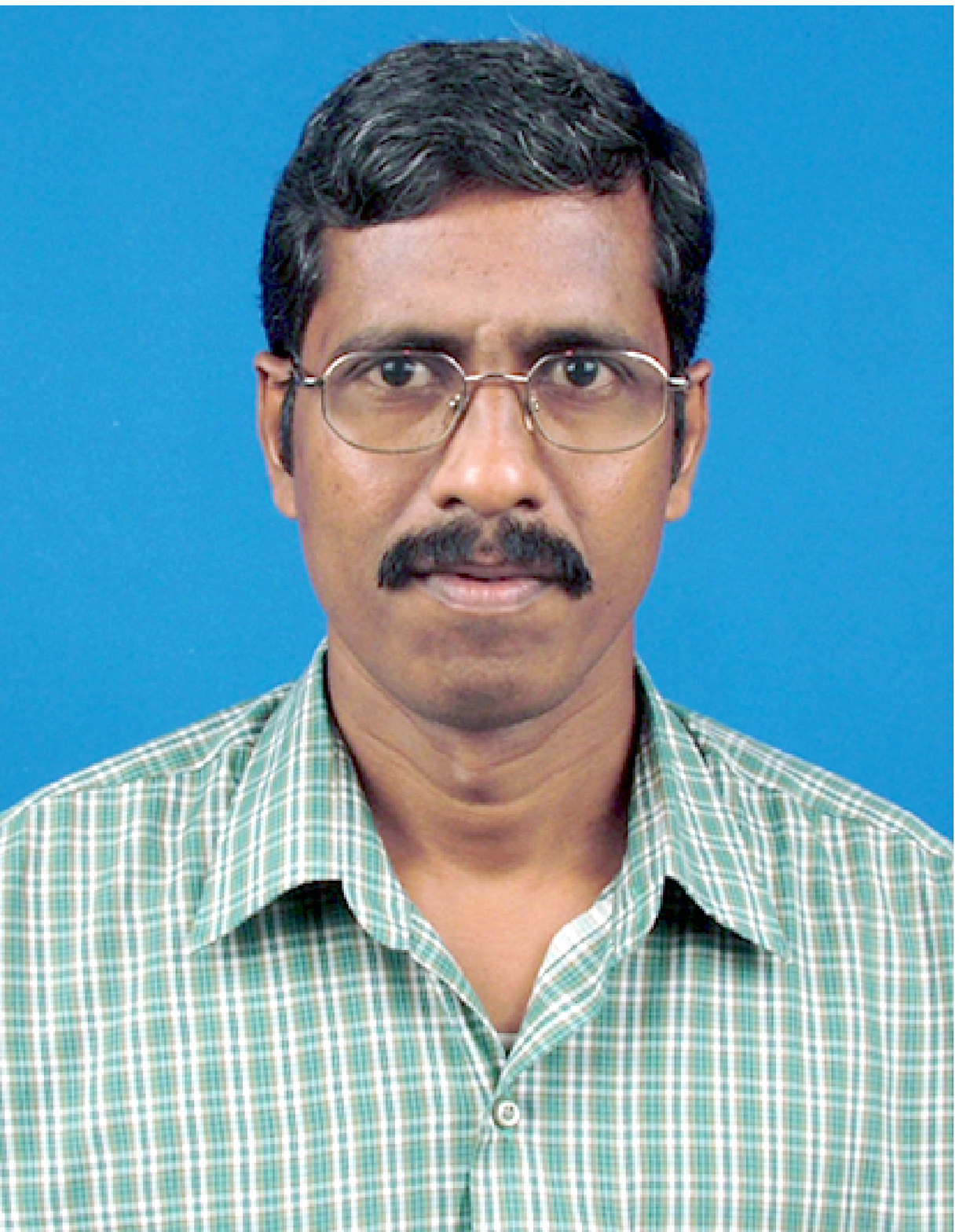}}
\begin{minipage}[b][1in][c]{1.8in} 
{\centering{ \bf{A Sreekumar}} received his MTech Degree in Computer Science and Engineering from IIT Madras, in 1992 and Ph.D in Cryptography from Cochin University of Science and Technology, in 2010. He joined as a Lecturer in the Department  of Computer Applications, CUSA-} \\\\
\end{minipage}
 T, in 1994 and currently he is working as an Associate Professor. He had more than 20 years of teaching experience. His research interest includes Cryptography, Secret Sharing Schemes and Number Theory.
\small\balance
\end{document}